\documentclass[12pt]{article}

\usepackage{graphicx}

\newcommand{\mat}{\left ( \begin{array}}
\newcommand{\emat}{\end{array} \right )}
\newcommand{\vect}{\left ( \begin{array}{c}}
\newcommand{\evect}{\end{array} \right )}

\begin{document}
%\titlerunning{Title running}
\begin{center}
{\Large\bf \boldmath Mesons and diquarks in the CFL phase of dense
quark matter} %<== title (bold face, capitalize)

\vspace*{6mm}
{D. Ebert$^{a,b}$ and K.G. Klimenko$^{c}$ }\\      %<== authors
{\small \it $^a$ Institute of Physics, Humboldt-University Berlin,
12489 Berlin, Germany \\      %<== institutions
 $^b$ Bogoliubov Laboratory of Theoretical Physics, JINR 141980
 Dubna, Russia \\          
			$^c$ IHEP and
 Dubna University (Protvino branch), 142281 Protvino, Russia}
\end{center}
%%%%%%%%%%%%%%%%%%%%%%%%%%%%%%%%%%%%%%%%%%%%%%%%%%%%%%%%%%%%%%%%%%%%
%   Please insert for me also my Dubna address:
%    Bogoliubov Laboratory of Theoretical Physics, JINR 141980 Dubna,
%Russia
%%%%%%%%%%%%%%%%%%%%%%%%%%%%%%%%%%%%%%%%%%%%%%%%%%%%%%%%%%%%%%%%%%%%

\vspace*{6mm}

% abstract
\begin{abstract}
The spectrum of  meson and diquark excitations of the color--flavor
locked (CFL) phase of dense quark matter is considered in the
framework of the Nambu -- Jona-Lasinio model. We have found that in
this phase all Nambu--Goldstone bosons are realized as scalar and
pseudoscalar diquarks. Other diquark excitations are resonances with
mass value around 230 MeV. Mesons are stable particles in the CFL
phase. Their masses vs chemical potential lie in the interval
300$\div$500 MeV.
\end{abstract}

\vspace*{6mm}

%Use this template to prepare your manuscript for Proceedings of the
%Conference.
%Please, send your contribution to \verb|spmtp08@theor.jinr.ru|.
%\textbf{Deadline
%for submission is September 15, 2008. Pagelimit is 8 pages for
%plenary (40 min) talk, 6 pages for 30 min talk, 3 pages for 20 min
%talk.}

%\section{Introduction}

It is well-known that at asymptotically high baryon densities the
ground state of massless three-flavor QCD corresponds to the
so-called color -- flavor locked (CFL) phase
\cite{alford1}.
One of the most noticeable differences between color
superconductivity phenomena with three and two quark species is that
the CFL effect is characterized by a hierarchy of energy scales.
Indeed, at the lowest scale of this phase lie NG bosons, which
dominate in all physical processes with energies
smaller than the superconducting gap $\Delta$. Evident contributors
at higher energy scales are quark quasiparticles, which in the CFL
phase have an energy greater than $\Delta$. However, up to now we
know much less about other excitations, whose energy and mass are of
the order of $\Delta$ in magnitude. Among these particles are
ordinary scalar and pseudoscalar mesons, massive diquarks etc, i.e.
particles which might play an essencial role in dynamical processes
of the CFL phase.
Here we are going to discuss just this type of excitations of the
CFL ground state, i.e. mesons and massive diquarks, in the framework
of the massless three-flavor NJL model with Lagrangian
\begin{eqnarray}
&&  L=\bar q\Big [\gamma^\nu i\partial_\nu+  \mu\gamma^0\Big ]q+
G_1\sum_{a=0}^8\Big [(\bar
q\tau_aq)^2+(\bar qi\gamma^5\tau_a q)^2\Big ]+\nonumber\\
&&G_2\!\!\!\sum_{A=2,5,7}\sum_{A'=2,5,7}\Big\{ [\bar
q^Ci\gamma^5\tau_A\lambda_{A'}q] [\bar qi\gamma^5\tau_A\lambda_{A'}
q^C] +[\bar q^C\tau_A\lambda_{A'}q] [\bar q\tau_A\lambda_{A'}
q^C]\Big\}.
   \label{1}
\end{eqnarray}
In (\ref{1}), $\mu\geq 0$ is the quark number chemical potential
which is the same for all quark flavors, $q^C=C\bar q^t$, $\bar
q^C=q^t C$ are charge-conjugated spinors, and $C=i\gamma^2\gamma^0$
is the charge conjugation matrix (the symbol $t$ denotes the
transposition operation). The quark field $q$ is a flavor and color
triplet as well as a four-component Dirac spinor. Furthermore, we
use the notations $\tau_a,\lambda_a$ for Gell-Mann matrices in the
flavor and color space, respectively ($a=1, ...,8)$; $\tau_0
=\sqrt{\frac{2}{3}}\cdot\bf 1_f$ is proportional to the unit matrix
in the flavor space. Clearly, the Lagrangian (\ref{1}) as a whole is
invariant under transformations from the color group SU(3)$_c$. In
addition, it is symmetric under the chiral group
SU(3)$_L\times$SU(3)$_R$ as well as under the baryon-number
conservation group U(1)$_B$ and the axial group U(1)$_A$.
\footnote{In a more realistic case, the additional `t Hooft
six-quark interaction term should be taken into account in order to
break the axial U(1)$_A$ symmetry.} In all numerical calculations
below, we used the following values of the model parameters:
$\Lambda=602.3$ MeV, $G_1\Lambda^2=2.319$ and $G_2=3G_1/4$, where
$\Lambda$ is an ultraviolet cutoff parameter in the
three-dimensional momentum space.

Introducing collective scalar $\sigma_a (x),\Delta^{s}_{AA'}(x)$ and
pseudoscalar $\pi_a (x),\Delta^{p}_{AA'}(x)$  fields,
\begin{eqnarray}
&&\sigma_a(x)=-2G_1(\bar q\tau_aq),~~~\Delta^s_{AA'}(x)=-2G_2(\bar
q^Ci\gamma^5\tau_A\lambda_{A'}q), \nonumber\\&&\pi_a(x)=-2G_1(\bar
qi\gamma^5\tau_a q),~~~ \Delta^p_{AA'}(x)=-2G_2(\bar
q^C\tau_A\lambda_{A'}q), \label{3}
\end{eqnarray}
($a=0,1,..,8; A,A'=2,5,7$) and then integrating out quark fields
from the theory, it is possible to obtain the following generating
functional of the two-point one-particle irreducible (1PI) Green
functions of the mesons and diquarks in the CFL phase
\begin{eqnarray}
&&{\cal S}_{\rm {eff}}
(\sigma_a,\pi_a,\Delta^{s,p}_{AA'},\Delta^{s,p*}_{AA'})
  = -\int d^4x\left[\frac{\sigma^2_a+\pi^2_a}{4G_1}+
\frac{\Delta^s_{AA'}\Delta^{s*}_{AA'}+\Delta^p_{AA'}
\Delta^{p*}_{AA'}}{4G_2}\right]+\nonumber\\
&&~~~~~~~~~~~~~~~~~~~~~~\frac i4{\rm Tr}\left\{S_0\left
(\begin{array}{cc}
\Sigma~, & K\\
  K^*~, & \Sigma^t
\end{array}\right )S_0\left (\begin{array}{cc}
\Sigma~, & K\\
  K^*~, & \Sigma^t
\end{array}\right )\right\},
   \label{12}
\end{eqnarray}
where
\begin{eqnarray}
&&%D^+=i\gamma^\nu\partial_\nu+\mu\gamma^0-\Sigma,~~~~~~~
\Sigma=\tau_a\sigma_a+ i\gamma^5\pi_a\tau_a,~~~~~~~
K=(\Delta^p_{AA'}+i\Delta^s_{AA'}\gamma^5)\tau_A\lambda_{A'},
\nonumber\\&&
%D^-=i\gamma^\nu\partial_\nu-\mu\gamma^0-\Sigma^t,~~~~~~
\Sigma^t=\tau_a^t\sigma_a+ i\gamma^5\pi_a\tau^t_a,~~~~~~
K^*=(\Delta^{p*}_{AA'}+i\Delta^{s*}_{AA'}\gamma^5)\tau_A
\lambda_{A'}, \label{6}
\end{eqnarray}
$S_0$ is the Nambu -- Gorkov representation for the quark propagator
in this phase,
\begin{equation}
S^{-1}_0=\left (\begin{array}{cc}
i\gamma^\nu\partial_\nu+\mu\gamma^0~, &
-i\Delta\gamma^5(\tau_2\lambda_{2}
+\tau_5\lambda_{5}+\tau_7\lambda_{7})\\
  -i\Delta\gamma^5(\tau_2\lambda_{2}
+\tau_5\lambda_{5}+\tau_7\lambda_{7})~, &
i\gamma^\nu\partial_\nu-\mu\gamma^0
\end{array}\right ),
\label{9}
\end{equation}
and $\Delta$ is the gap parameter. Using the expression (\ref{12}),
one can find the 1PI Green functions for mesons and diquarks in the
CFL phase, namely
\begin{eqnarray}
&& \Gamma_{XY}(x-y)=-\frac{\delta^2{\cal S}^{(2)}_{\rm eff}}{\delta
Y(y)\delta X(x)},
  \label{17}
\end{eqnarray}
where $X(x),Y(x)=\sigma_a(x),\pi_b(x),\Delta^{s,p}_{AA'}(x),
\Delta^{s,p*}_{BB'}(x)$. In total, the Green functions (\ref{17})
form a 54$\times$54 matrix which is, fortunately, a reducible one.
It is well known that in the rest frame of the momentum space
representation, i.e. at $p=(p_0,0,0,0)$, the meson and diquark
masses are the zeros of the determinant of this matrix. So, after
tedious both analytical and numerical calculations (the details are
presented in \cite{ek}) we have obtained the following results on
the mass spectrum of the bosonic excitations of the CFL phase.

In the Figs 1,2 the mass behavior for the scalar and pseudoscalar
mesons is presented in the CFL phase. It is easily seen that i)
there is a mass splitting between octet and singlet of mesons, ii)
the singlet mass is smaller than octet one for scalar mesons, but
for pseudoscalar mesons the situation is inverse, iii) in the CFL
phase the meson masses are greater than 300 MeV.

In the diquark sector we have found 18 Nambu-Goldstone bosons.
Moreover, there are a scalar octet and singlet as well as
pseudoscalar octet and singlet of nontrivial diquark excitations of
the CFL phase. The diquarks from scalar and pseudoscalar octets are
resonances with mass around 230 MeV. The properties of diquarks in
the CFL phase were also considered earlier in the papers
\cite{ek,ruggieri}.

For comparison, let us mention the existence of
nontrivial excitations of the
color superconducting (2SC) phase of quark matter with two quark
flavors. In this phase the masses of $\sigma$ and $\pi$ mesons lie
in the same interval 300$\div$500 MeV as the meson masses in the
CFL phase. But the scalar diquark is a very heavy resonance with
mass $\sim$ 1100 MeV \cite{bekvy}. If the electric charge
neutrality constraint is imposed, then in the 2SC phase diquark is a
stable particle with mass $\sim$200 MeV \cite{eky2}.

%%%%%%%%%% Acknowledgement
We thank V.L. Yudichev for the fruitful cooperation over many years.
One of us (D.E.) thanks A.E. Dorokhov and M.K. Volkov for useful
discussions
and the Bogoliubov Laboratory of Theoretical Physics for kind
hospitality.
\begin{figure}
%----figure 1
  \includegraphics[width=0.45\textwidth]{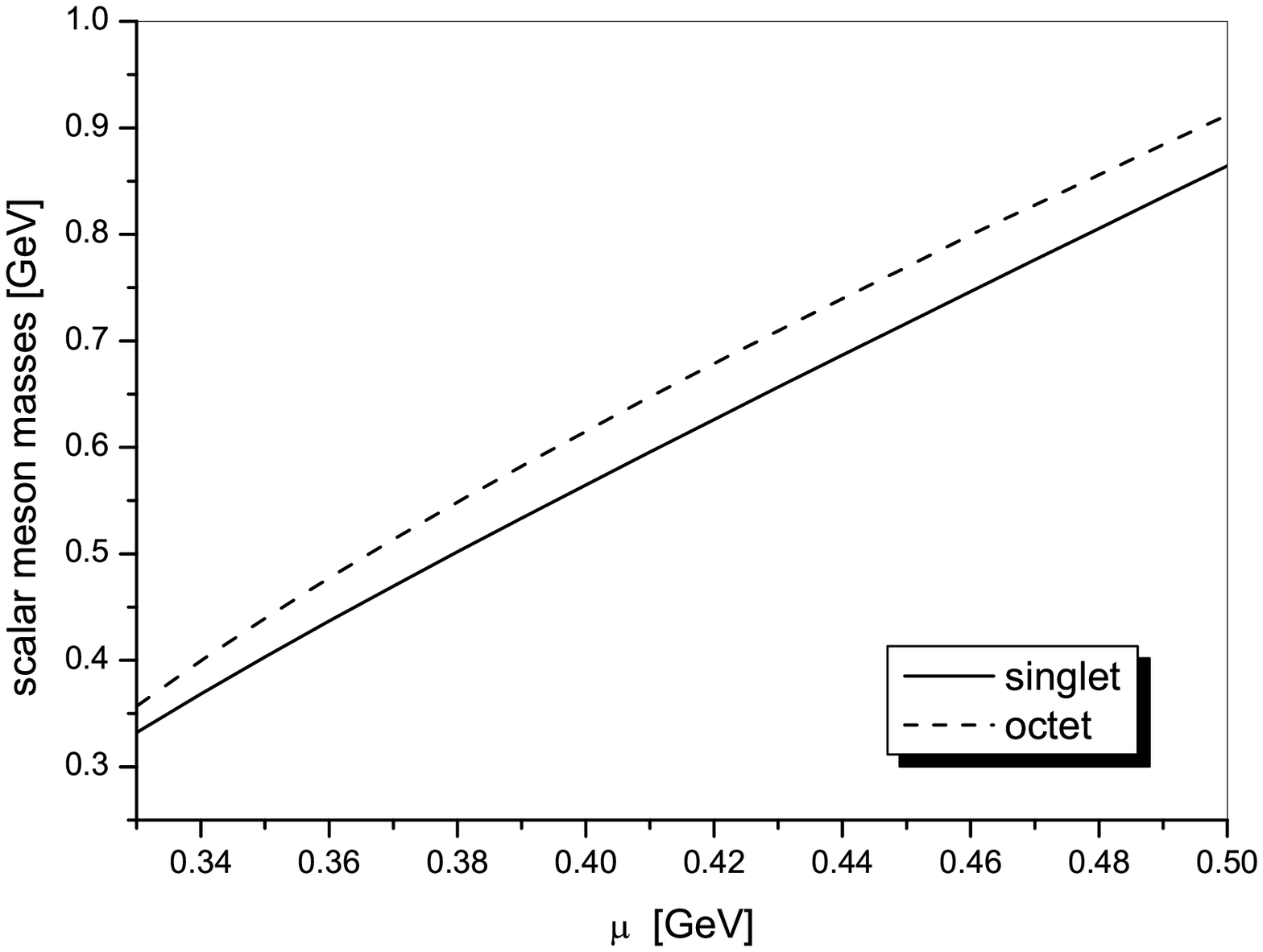}
  \hfill
  \includegraphics[width=0.45\textwidth]{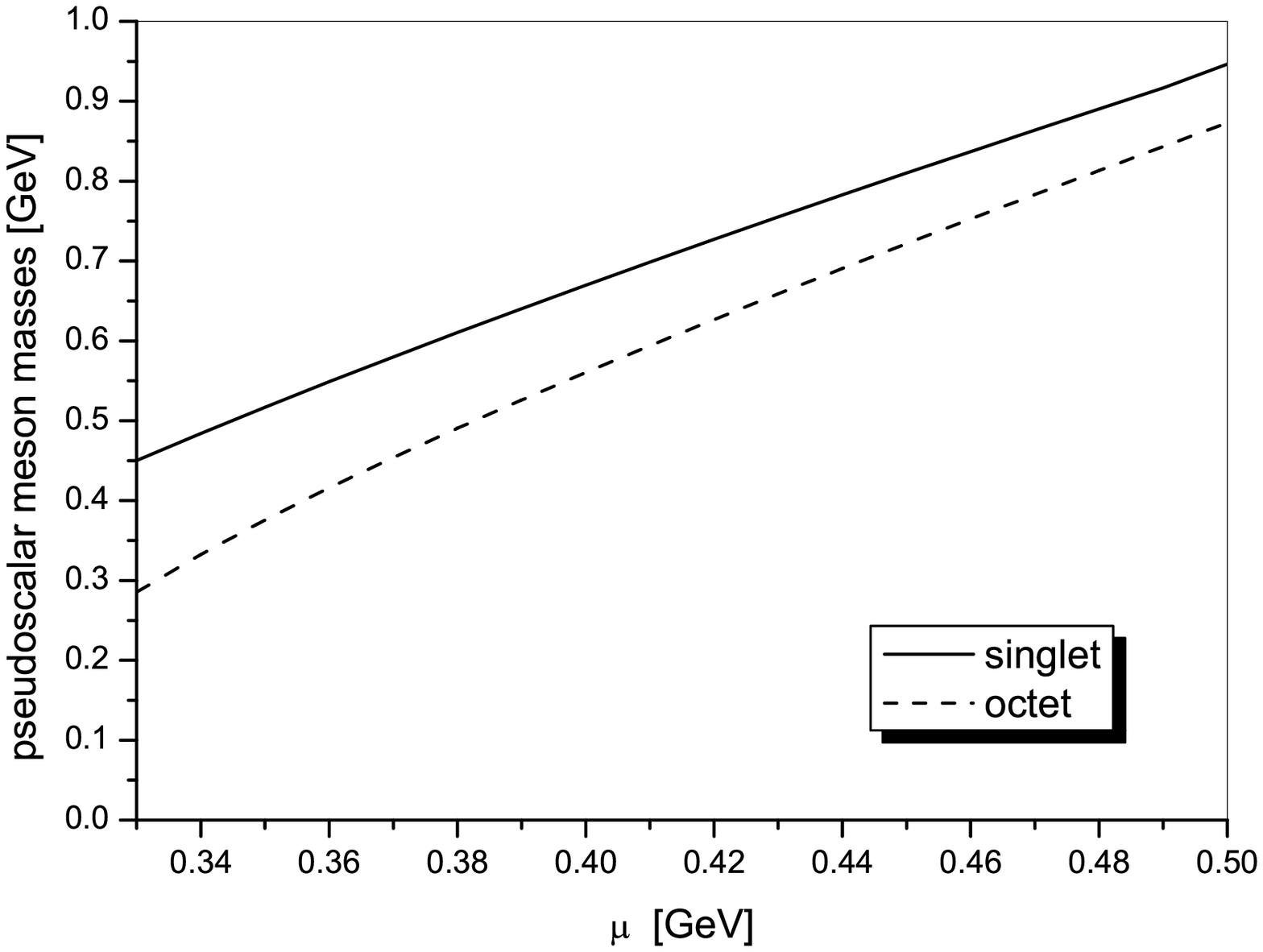}\\
\parbox[t]{0.45\textwidth}{
  \caption{The behavior of the scalar meson masses vs $\mu$ in the
CFL phase.} \label{fig:1} } \hfill
\parbox[t]{0.45\textwidth}{
\caption{The behavior of the pseudoscalar meson masses in
  the CFL phase.}
\label{fig:2} }
\end{figure}


\begin{thebibliography}{99}\itemsep -1mm
%\bibitem{tmpref} T.M.~Plate {\em et al}., Journal {\bf 31}, 3
%(2007).
\bibitem{alford1}
M. Alford, K. Rajagopal, and F. Wilczek, Nucl. Phys. B {\bf 537},
443 (1999);
M. Alford, J. Berges, and K. Rajagopal, Nucl. Phys. B {\bf 558}, 219
(1999);
M. Alford, A. Schmitt, K. Rajagopal, T. Schafer, arXiv:0709.4635.

\bibitem{ek}
D.~Ebert and K.G.~Klimenko, Phys.\ Rev. D {\bf 75}, 045005 (2007).
%%CITATION = HEP-PH 0611385;%%
D.~Ebert, K.G.~Klimenko, and V.L.~Yudichev, Eur. Phys. J. C {\bf
53}, 65 (2008).
%%CITATION = EPHJA,C53,65;%%

\bibitem{ruggieri}
M. Ruggieri, JHEP {\bf 0707}, 031 (2007); V. Kleinhaus, M. Buballa,
D. Nickel, and M. Oertel, Phys.\ Rev. D {\bf 76}, 074024 (2007); T.
Brauner, Phys.\ Rev. D {\bf 77}, 096006 (2008); D. Zablocki, D.
Blaschke, and R. Anglani, arXiv:0805.2687.

\bibitem{bekvy}
D.~Blaschke {\em et al}., Phys.\ Rev. D {\bf 70}, 014006 (2004);
%%CITATION = HEP-PH 0403151;%%
%\bibitem{eky}
D.~Ebert, K.G.~Klimenko, and V.L.~Yudichev, Phys.\ Rev. C {\bf 72},
015201 (2005);
%%CITATION = HEP-PH 0412129;%%
Phys.\ Rev. D {\bf 72}, 056007 (2005);
%%CITATION = HEP-PH 0504218;%%
%\bibitem{klim}
D.~Ebert and K.G.~Klimenko, Theor. Math. Phys. {\bf 150}, 82 (2007).
%%CITATION = TMPHA,150,82;%%

\bibitem{eky2}
D.~Ebert, K.G.~Klimenko, and V.L.~Yudichev, Phys.\ Rev. D {\bf 75},
025024 (2007).
%%CITATION = HEP-PH 0608304;%%

\end{thebibliography}
\end{document}